\documentclass[%
twocolumn,
amsmath,amssymb,
aps, 
 nofootinbib,
]{revtex4-2}
\usepackage[utf8]{inputenc}
\usepackage[T1]{fontenc}
\usepackage{braket}
\usepackage{float}
\newcommand{\beginsupplement}{%
        \setcounter{table}{0}
        \renewcommand{\thetable}{S\arabic{table}}%
        \setcounter{figure}{0}
        \renewcommand{\thefigure}{S\arabic{figure}}%
     }
\usepackage{blindtext}
\usepackage[breaklinks]{hyperref}
\usepackage{algorithm}
\newfloat{algorithm}{t}{lop}
\usepackage{algpseudocode}
\usepackage{graphicx}
\newcommand{\argmin}{\operatornamewithlimits{argmin}}

\usepackage{xcolor}
\usepackage{dcolumn}
\usepackage{lineno}
\usepackage{placeins}

\usepackage{bm}

\begin{document}

\preprint{APS/123-QED}

\title{Approximating Excited States using Neural Networks}

\author{Yimeng Min}

\thanks{ min@cs.cornell.edu}

\affiliation{%
Department  of  Computer  Science\\
Cornell  University\\  Ithaca,  NY  14853,  United  States
}%

\begin{abstract}
Recently developed neural network-based wave function methods are capable of achieving state-of-the-art results for
finding the ground state in real space.
In this work, a neural network-based  method is used to  compute excited states.
We train our network via  variational principle, along a further penalty term that imposes the orthogonality with lower-energy eigenfunctions.
As a demonstration of the effectiveness of this approach, results from numerical calculations for one-dimensional and two-dimensional harmonic oscillators are presented.

%

\end{abstract}

\maketitle

\section{Introduction}
Neural networks (NNs), inspired by biological processes of neurons in a brain, have become pioneering methods in pattern recognition and machine learning in recent years \cite{schmidhuber2015deep}\cite{lecun2015deep}. 
Recently, NN-based models have shown great promise in physical science, such as identifying phases and phase transitions, designing quantum experiments and simulating the ground state of a quantum system \cite{carrasquilla2017machine}\cite{melnikov2018active}\cite{carleo2017solving}.
These models can be categorized into two different types:  data-driven and data-free methods.

Most applications of NN-based models to physical sciences use data-driven methods, where the models are trained using external data (training data) and the predictions are performed separately on test data  \cite{schutt2019unifying}\cite{rupp2012fast}. The inputs can be configurations sampled with Monte Carlo \cite{carrasquilla2017machine}, datasets compiled from existing computational methods like density functional theory (DFT) \cite{schutt2019unifying}\cite{gilmer2017neural} or exact solutions on a lattice \cite{mills2017deep}. For date-free NN-based models, motivated by the fact that NNs are universal function approximators \cite{hornik1989multilayer},  people tend to use NNs as wave function ansatzes and minimize the energy expectation value. This is known as the variational method: if the NN-based wave function is close enough to the true ground state, so is the minimum energy expectation value.

Recently, Carleo and Troyer propose NN-based representations for discrete spin lattice systems and train their networks via the variational principle \cite{carleo2017solving}. 
Besides spin lattice systems,  NN-based representations are also demonstrated to significantly reduce the relative energy error of variational ground state in real space \cite{hermann2020deep}\cite{pfau2020ab}\cite{teng2018machine}.

So far,  these NN-based variational methods have focused on the ground-state properties only. For spin lattice systems, the  excited states provide information such as  ground state degeneracy, size of the excitation gap, and low-lying dispersion of excitations \cite{choo2018symmetries}\cite{nomura2020machine}\cite{vieijra2020restricted}. Different types of NN-based variational approaches have been utilized to model the excited states of spin systems, for example, Choo $et$ $al.$  take advantage of Abelian spatial symmetries and  orthogonality between wave function with respect to the ground state  to obtain  the  energy  gap  between  the  ground  state  and  the  first  excited state \cite{choo2018symmetries}, Nomura   further extends the this method using a  smaller number of variational parameters \cite{nomura2020machine}, 
 Vieijra $et$ $al.$ construct a variational wave function that transforms  as  an  irreducible  representation  of  SU(2) and provides direct access to the construction of excited states in spin systems \cite{vieijra2020restricted}. 

Though researchers have been trying to use NN-based variational methods for simulating the excited states in spin systems, in real space, data-driven methods are dominant approaches to approximate the excited states \cite{mills2017deep}\cite{westermayr2020machine}.  These NN-based  data-driven approaches avoid the solution of the Schr\"{o}dinger equation at the price of requiring preexisting solutions such as spectrum datasets \cite{kiyohara2020learning} and photodynamics datasets \cite{westermayr2020combining}. Furthermore, data-driven methods  highly depend on the generalization ability of the NN structures and the size of training datasets.
For improving the  simulation accuracy (generalization ability), researchers need to simulate a very large dataset or design   complicated and handcrafted NN structures to encode the input information. However, the first strategy is very challenging as the dataset can subject to a heavy-tailed distribution, which means it is unrealistic to cover all the instances. For the latter one, 
it is also very difficult
to determine the key parameters which control the approximation accuracy because these complicated structures typically involve a huge number of parameters.

In this paper, we propose a NN-based penalty method to simulate single particle excited states in real space using variational method, where no external data are required. Similar techniques are used in density matrix renormalization group (DMRG) calculations and variational Monte Carlo (VMC) \cite{stoudenmire2012studying}\cite{pathak2021excited}, but to our knowledge have not been applied in the NN-based wave function ansatz context. 

We add orthogonal penalty with lower-energy eigenfunctions  into the variational loss. This additional penalty enables us to efficiently approximate the excited states.
We construct the wave function using a fully connected neural network with one hidden layer, which is among one of the simplest NN structures. We further study how the parameters of neural networks can affect the approximation accuracy. 

This paper is organized as follows. In section~\ref{sec:1}, we introduce network structures  and review variational theory , we show the orthogonal penalty can lead to the rearrangement of the eigenvalues. Section~\ref{sec:2} presents our results on one-dimensional and two-dimensional quantum harmonic oscillators. In section~\ref{sec:3} and \ref{sec:4}, we conclude our method and perform error analysis, we further compare the number of parameters used in our model with  NN-based data-driven methods.
\nopagebreak
\section{Methods\label{sec:1}}
\subsection{Neural network representation}
\begin{figure}[H]
         \centering
         \includegraphics[width=0.8\columnwidth]{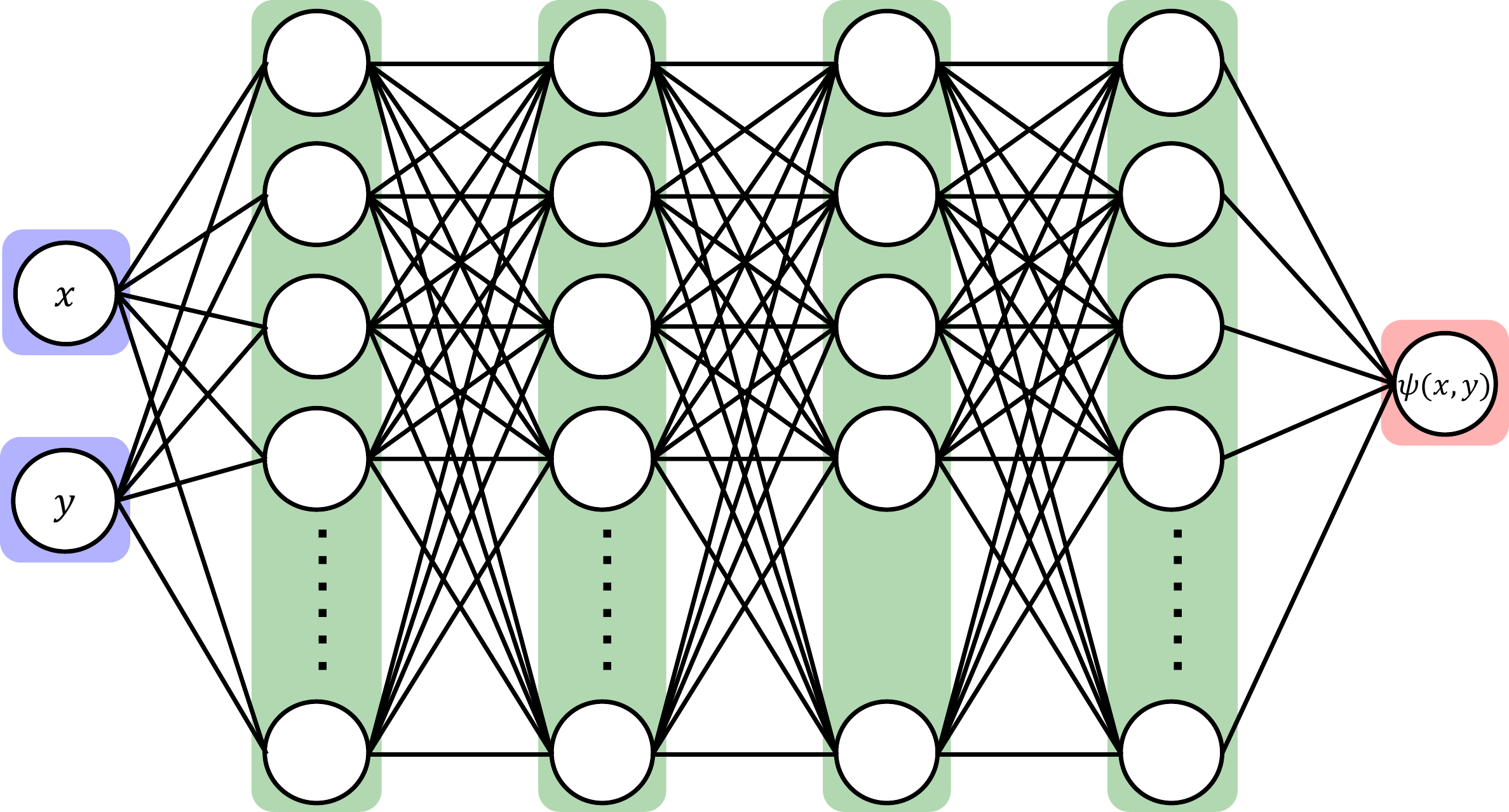} 
         \caption{Neural network architecture for 2D cases. The network takes the coordinates as the input and the output is the  wave function.}
         \label{fig:nnske}
\end{figure}

We start by building a neural network to represent the real part of the wave function $\ket{\psi(\boldsymbol{\vec x})}$:\footnote{We assume the phase of wave function is zero.}
\begin{equation}
    \ket{\psi(\boldsymbol{\vec x})} = f_{nn}(\boldsymbol{\theta})(\boldsymbol{\vec x}),
\end{equation}
where $\boldsymbol{\vec x} \in \mathbb{R}^d$ is the coordinate, $f_{nn}(\boldsymbol{\theta}): \mathbb{R}^d\rightarrow\mathbb{R}$ is the neural network function which maps the input coordinate to the wave function. The input consists of $d$ neurons, followed by $l$ hidden layers and each hidden layer has $\sigma$ neurons.  The hyperbolic tangent function ($Tanh$) is used as a nonlinear activation after each hidden layer. The output layer has connections to all activated neurons in the previous layer. Let $\boldsymbol{\theta}$ represent the parameters in the neural network. Figure~\ref{fig:nnske}  illustrates the neural network structure we used for two-dimensional quantum harmonic oscillators .
In our simulation, the plane is discretized into $M_x$ by $M_y$ points and the expectation value of the Hamiltonian is evaluated on these coordinates.

\subsection{Variational method}
Variational computations have mostly been used to investigate the ground-state properties of various systems. 
Consider a Hamiltonian operator $\hat{\boldsymbol{H}} \in \mathbb{R}^{N\times N}$ whose eigenstates are $\ket{\psi_0}$, $\ket{\psi_1}$, $\ket{\psi_2}$...$\ket{\psi_{N-1}}$ and whose eigenvalues are $e_0\leq e_1\leq e_2...\leq e_{N-1}$. For approximating the ground state, we minimize the  energy expectation
\begin{equation}\label{eq:ground}
        \boldsymbol{E_0}(\boldsymbol{\theta}) = \frac{\bra{\psi(\boldsymbol{\theta})}\hat{\boldsymbol{H}}\ket{\psi(\boldsymbol{\theta})}}{\braket{\psi(\boldsymbol{\theta})|\psi(\boldsymbol{\theta})}} .
\end{equation}
Here $\ket{\psi(\boldsymbol{\theta})}$ is the neural network representation of the wave function and $\boldsymbol{
\theta}$ represents the parameters to be optimized.  For the simplicity of notation, let $\ket{\psi(\boldsymbol{\theta})}$ denote the normalized wave function $\frac{\ket{\psi(\boldsymbol{\theta})}}{\sqrt{\braket{\psi(\boldsymbol{\theta})|\psi(\boldsymbol{\theta})}}}$, and equation~\ref{eq:ground} becomes:
\begin{equation}\label{eq:normground}
        \boldsymbol{E_0}(\boldsymbol{\theta}) = \bra{\psi(\boldsymbol{\theta})}\hat{\boldsymbol{H}}\ket{\psi(\boldsymbol{\theta})}.
\end{equation}
We then optimize $\boldsymbol{\theta}$  using variational principle,
\begin{equation}
    \tilde{\boldsymbol{\theta}} = \argmin_{\boldsymbol{\theta}}\boldsymbol{E_0}(\boldsymbol{\theta}) =  \argmin_{\boldsymbol{\theta}} \bra{\psi(\boldsymbol{\theta})}\hat{\boldsymbol{H}}\ket{\psi(\boldsymbol{\theta})}.
\end{equation}
Let $\ket{\boldsymbol{\tilde{\psi_0}}}$ denote the corresponding wave function $\ket{\psi(\tilde{\boldsymbol{\theta}})}$. $\ket{\boldsymbol{\tilde{\psi_0}}}$ approximates $\ket{\psi_0}$ because the minimum of the energy occurs when $\ket{\boldsymbol{\tilde{\psi_0}}}$ is the ground-state wave function of $\hat{\boldsymbol{H}}$.  
For approximating the first excited state,\footnote{
We assume the systems are not degenerate at ground energy level.} we minimize the following:
\begin{equation}\label{eq:1stex}
        \boldsymbol{E_1}(\boldsymbol{\theta}) = \bra{\psi(\boldsymbol{\theta})}\hat{\boldsymbol{H}}\ket{\psi(\boldsymbol{\theta})} + \lambda  \braket{\psi(\boldsymbol{\theta})|\boldsymbol{\tilde{\psi_0}}}^2.
\end{equation}
$\boldsymbol{E_1}(\boldsymbol{\theta})$ consists of two parts: the expectation value of the energy $\bra{\psi(\boldsymbol{\theta})}\hat{\boldsymbol{H}}\ket{\psi(\boldsymbol{\theta})}$ and the orthogonal penalty $\lambda  \braket{\psi(\boldsymbol{\theta})|\boldsymbol{\tilde{\psi_0}}}^2$, $\lambda > 0$.  
Our training minimizes the energy under the orthogonal constraint between $\ket{\psi(\boldsymbol{\theta})}$  and previously converged ground state $\ket{\boldsymbol{\tilde{\psi_0}}}$. Here, the orthogonal penalty parameter $\lambda$ governs the magnitude of the orthogonal penalty and determines the extent of how much the previous eigenvalue $e_0$ is lifted up. We rewrite equation~\ref{eq:1stex}:

\begin{align}
        & \boldsymbol{E_1}(\boldsymbol{\theta})  = \bra{\psi(\boldsymbol{\theta})}\hat{\boldsymbol{H}}\ket{\psi(\boldsymbol{\theta})} + \lambda  \braket{\boldsymbol{\tilde{\psi_0}}|\psi(\boldsymbol{\theta})}^2 \nonumber\\
        & =  \bra{\psi(\boldsymbol{\theta})}\big(\sum_{i=0}^{N-1} e_i \ket{\psi_i} \bra{\psi_i} + \ket{\boldsymbol{\tilde{\psi_0}}}\lambda\bra{\boldsymbol{\tilde{\psi_0}}}\big)\ket{\psi(\boldsymbol{\theta})}.
\end{align}
\label{eq:liftup1}

Since  $\ket{\boldsymbol{\tilde{\psi_0}}}$ can be regarded as the approximation of  $\ket{\psi_0}$,  the effective Hamiltonian $\hat{\boldsymbol{H}}_1 = \sum_{i=0}^{N-1} e_i \ket{\psi_i} \bra{\psi_i} + \lambda \ket{\boldsymbol{\tilde{\psi_0}}}\bra{\boldsymbol{\tilde{\psi_0}}}$ can be expressed in terms of the same basis set:
\begin{equation}
    \hat{\boldsymbol{H}}_1 = \sum_{i=1}^{N-1} e_i \ket{\psi_i} \bra{\psi_i} + (e_0+\lambda) \ket{\psi_0} \bra{\psi_0}.
\end{equation}
As long as $\lambda>e_1 - e_0$, then the lowest eigenvalue of $\hat{\boldsymbol{H}}_1$ becomes $e_1$.
By minimizing $\boldsymbol{E_1}(\boldsymbol{\theta})$, the variational method finds the new ground state  $\ket{\psi_1}$.

We can now generalize this penalty-based optimization scheme to $k$-$th$ excited state, let $p_i (\boldsymbol{\theta})$=$ \braket{\psi(\boldsymbol{\theta})|\boldsymbol{\tilde{\psi_i}}}$ represent the orthogonal penalty between the wave function $\ket{\psi(\boldsymbol{\theta})}$ and $i$-${th}$ excited state $\ket{\boldsymbol{\tilde{\psi_i}}}$. $\boldsymbol{p}_i(\boldsymbol{\theta}) =\big( p_0(\boldsymbol{\theta}),p_1(\boldsymbol{\theta}),p_2(\boldsymbol{\theta})...p_{i-1}(\boldsymbol{\theta})\big)$. 
For the $k$-${th}$ excited state, we minimize the following:
\begin{equation} \label{eq:nthex}
     \boldsymbol{E_k}(\boldsymbol{\theta}) = \bra{\psi(\boldsymbol{\theta})}\hat{\boldsymbol{H}}\ket{\psi(\boldsymbol{\theta})} + \lambda  ||\boldsymbol{p}_{k}(\boldsymbol{\theta})||_2^2.
\end{equation}
Let $\hat{\boldsymbol{H}}_k$ be the effective Hamiltonian for $k$-$th$ excited state. The orthogonal penalty lifts up the eigenvalues of $\ket{\psi_0}, \ket{\psi_1}... \ket{\psi_{k-1}}$ and one can express $\hat{\boldsymbol{H}}_k$ as:
\begin{equation}
    \hat{\boldsymbol{H}}_k = \sum_{i=k}^{N-1} e_i \ket{\psi_i} \bra{\psi_i} + \sum_{i=0}^{k-1} (e_i+\lambda) \ket{\psi_i} \bra{\psi_i},
\end{equation}
when $\lambda > e_k - e_0$, the orthogonal penalty will rearrange the eigenvalues. In other words, $\ket{\psi_k}$ becomes the new ground state because $ e_k< e_{0}+\lambda\leq e_{1}+\lambda\leq e_{2}+\lambda\leq e_{k-1}+\lambda$.  Thus, the neural network will no longer converge to previous states using this penalty-based optimization scheme.
In order to approximate $k$-$th$ excited state, $\lambda$ should be greater than $e_k - e_0$. This iterative training method enables us to find the $k$-$th$ excited states. However, the gap between eigenvalues can be unknown under some circumstances and we are unable to determine the threshold value $e_k - e_0$. In this condition, the penalty method still works but takes more iterations. Take the first excited state as an example, let $\Delta$ denote the gap between $e_1$ and $e_0$, when $\lambda$ is smaller than $\Delta$, it will take at least $\lceil \frac{\Delta}{\lambda} \rceil$ iterations to approximate $e_1$ and $\ket{\psi_1}$.  In our simulation, we set $\lambda = 1000$ to ensure $e_0$, $e_1$, ...,  $e_{k-1}$ are lifted up.

\subsection{Optimizing the wave function}
In this paper, we simulate the eigenstates of one-dimensional and two-dimensional  harmonic oscillators.
In one-dimensional condition, the $x$-axis is discretized into $M_x$ points ranging from $-L
_x$ to $L_x$, The neural network takes the discrete  coordinate as the input and the output is the wave function.  The algorithm is shown in Algorithm~\ref{alg:1}. The Hamiltonian is
\begin{equation*}
    \hat{\boldsymbol{H}} = -\frac{\hbar^2}{2m}\hat{\boldsymbol{\nabla}}^2 + \frac{m^2 \omega^2}{2} \hat{\boldsymbol{x}}^2,
\end{equation*}
and the eigenstates are
\begin{equation*}
    \psi_n(x) = \frac{1}{\sqrt{2^n n!}}(\frac{m \omega}{\pi \hbar})^{1/4} e^{-\frac{-m\omega x^2 }{2\hbar}} \mathcal{H}_n(\sqrt{\frac{m\omega}{\hbar}}x),
\end{equation*}
where $\mathcal{H}_n$ represents $n$-${th}$ Hermite polynomial. In our simulation, we take $\hbar=m=\omega = 1$. 
\begin{algorithm}[H]
	\caption{Computing the $k$-${th}$ excited states  (1D)}
	\label{alg:1}
\begin{algorithmic}[1]
	\State \textbf{Input}: A neural network function $f_{\theta}$; $M_x$ grid points on $x$-axis \{$x_1$...$x_{M_x}$\}; orthogonal penalty parameter $\lambda$.
	\State \textbf{Output}: $k$-${th}$ excited state $\ket{\psi_k}$ = \big($f_{\theta}(x_1)$ ... $f_{\theta}(x_{M_x})$\big)
\For{$i$ in $0,1,...,k$} 
    \State $\ket{\psi_i(\boldsymbol{\theta})} \leftarrow  \frac{\ket{\psi_i(\boldsymbol{\theta})}}{\sqrt{\braket{\psi_i(\boldsymbol{\theta})|\psi_i(\boldsymbol{\theta})}}} $
    \State $\boldsymbol{p}_i(\boldsymbol{\theta}) = \{\}$
        \For{$j$ in $0,1,...,i-1$} \State append $\braket{\psi_i(\boldsymbol{\theta})|\boldsymbol{\tilde{\psi_j}}}$ to $\boldsymbol{p}_i(\boldsymbol{\theta}) $
        \EndFor
    \State  $\boldsymbol{\tilde{\theta}} \leftarrow  \argmin \{\bra{\psi_i(\boldsymbol{\theta})}\hat{\boldsymbol{H}} \ket{\psi_i(\boldsymbol{\theta})}$ + $\lambda$  $||\boldsymbol{p}_i(\boldsymbol{\theta})||_2^2 \}$  
    \State save the converged $\ket{\psi_i(\boldsymbol{\tilde{\theta}})}$ as $\ket{\boldsymbol{\tilde{\psi_i}}}$
\EndFor
\State \Return $\ket{\boldsymbol{\tilde{\psi_k}}}$
\end{algorithmic} 
\end{algorithm}
The Hamiltonian of two-dimensional quantum harmonic oscillator is:
\begin{equation*}
    \hat{\boldsymbol{H}} = -\frac{1}{2}\hat{\boldsymbol{\nabla}}^2 + \frac{1}{2} (\hat{\boldsymbol{x}}^2 + \hat{\boldsymbol{y}}^2),
\end{equation*}
and the eigenstates are:
\begin{equation*}
    \psi_{n}(x,y) = \psi_{n_x}(x) \psi_{n_y}(y), 
\end{equation*}
where $\psi_{n_x}$ is the $n_x$-$th$  one-dimensional eigenstate  along $x$ direction and $\psi_{n_y}$ is the $n_y$-$th$  one-dimensional eigenstate  along $y$ direction.

Similar to one-dimensional condition, we first mesh the $xy$ plane using $M_x\times M_y$ points ranging from  $(-L_x,-L_y)$ to $(L_x,L_y)$. The neural network takes the discrete coordinate $(x_i,y_j)$ as the input and the output is $\psi(x_i,y_j)$. 
The kinetic energy operator is
\begin{equation}\label{eq:2Dhar}
    -\frac{1}{2}\hat{\boldsymbol{\nabla}}^2=  -\frac{1}{2} (\widehat{\boldsymbol{L_{xx}}} \otimes \hat{\boldsymbol{I}} + \hat{\boldsymbol{I}} \otimes  \widehat{\boldsymbol{L_{yy}}  } ),
\end{equation}
where $ \widehat{\boldsymbol{L_{xx}}}$ and $ \widehat{\boldsymbol{L_{yy}}}$ are one-dimensional discrete Laplacian operators in the $x$ and $y$ direction, respectively.

\section{Results\label{sec:2}}
\subsection{One-dimensional quantum harmonic oscillator}
\begin{figure}[H]
     \centering
     \begin{minipage}{0.5\columnwidth}
         \centering
         \includegraphics[width=1.0\columnwidth]{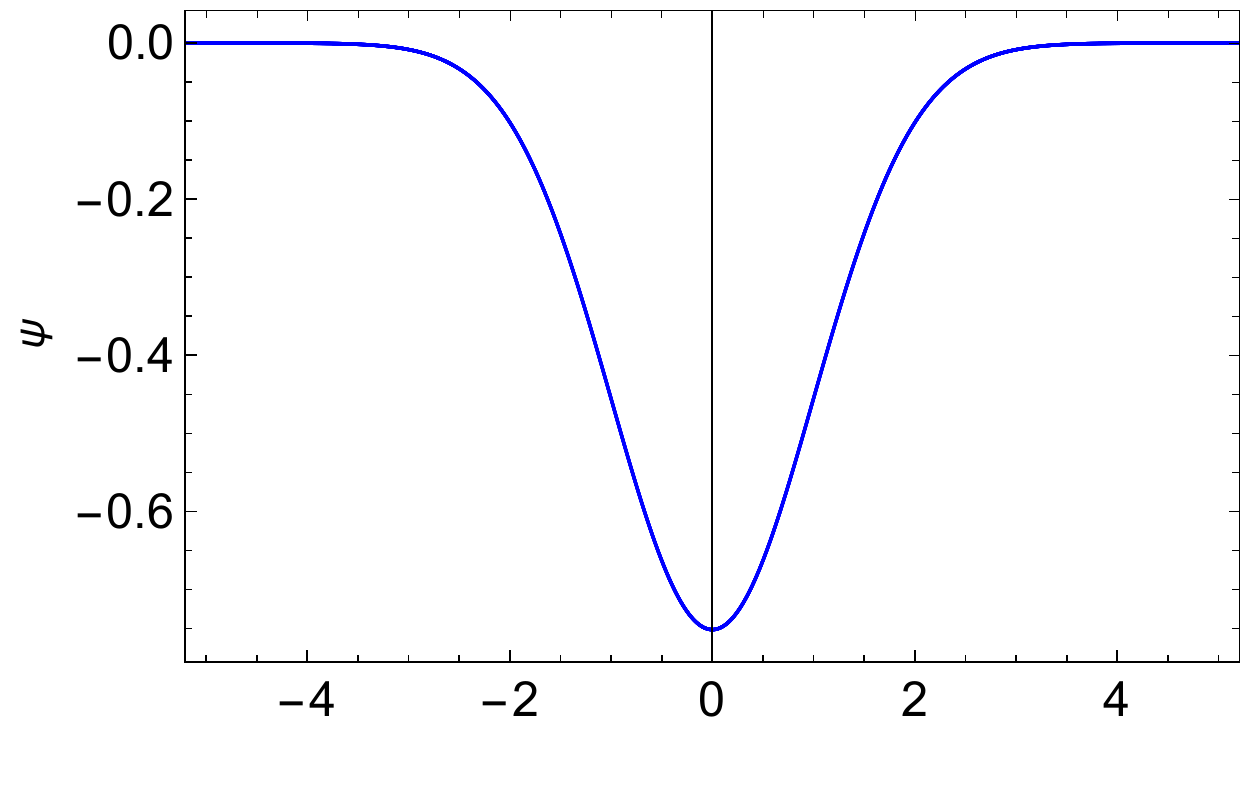} 
     \end{minipage}\hfill
     \begin{minipage}{0.5\columnwidth}
         \centering
         \includegraphics[width=1.0\columnwidth]{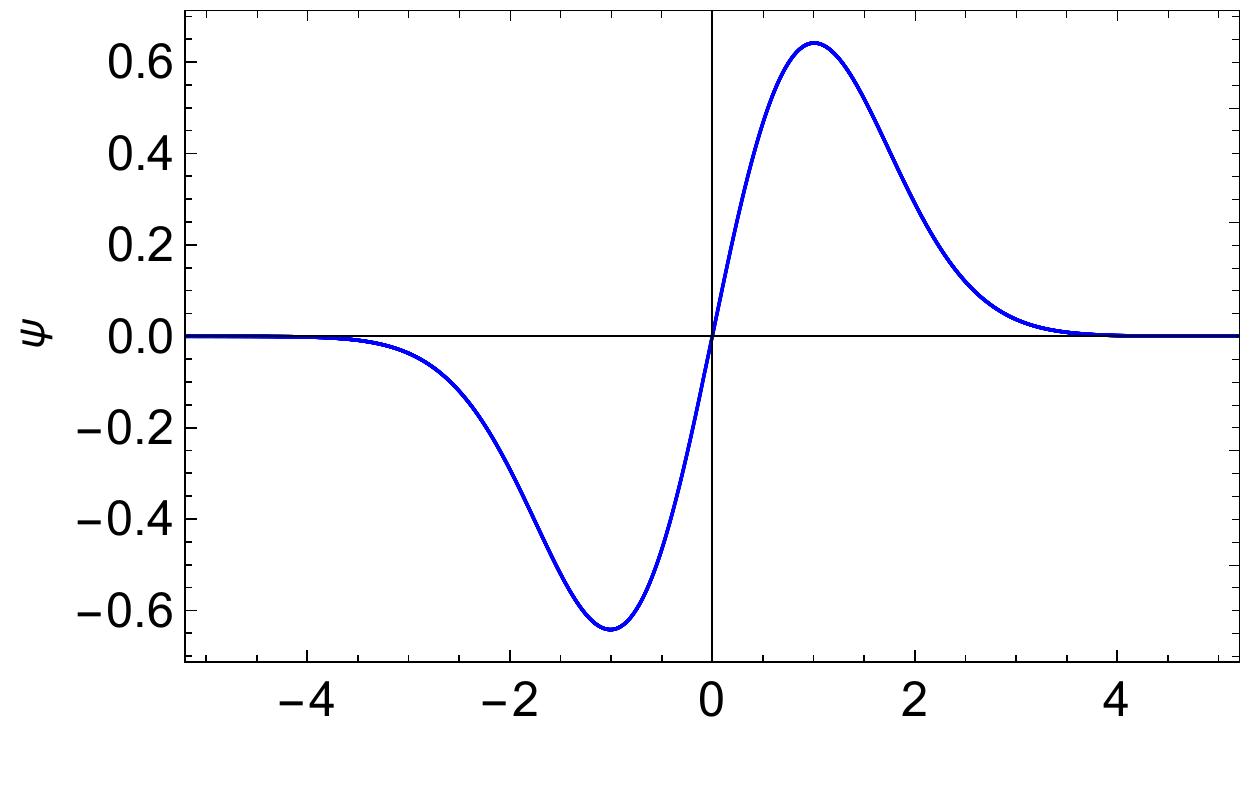} 
     \end{minipage}
     \caption{Predicted $\ket{\psi}$ of one-dimensional harmonic oscillator. Left: ground state; right: first excited state.}\label{fig:1D01}
 \end{figure}

\begin{figure}[H]
     \centering
     \begin{minipage}{0.5\columnwidth}
         \centering
         \includegraphics[width=1.0\columnwidth]{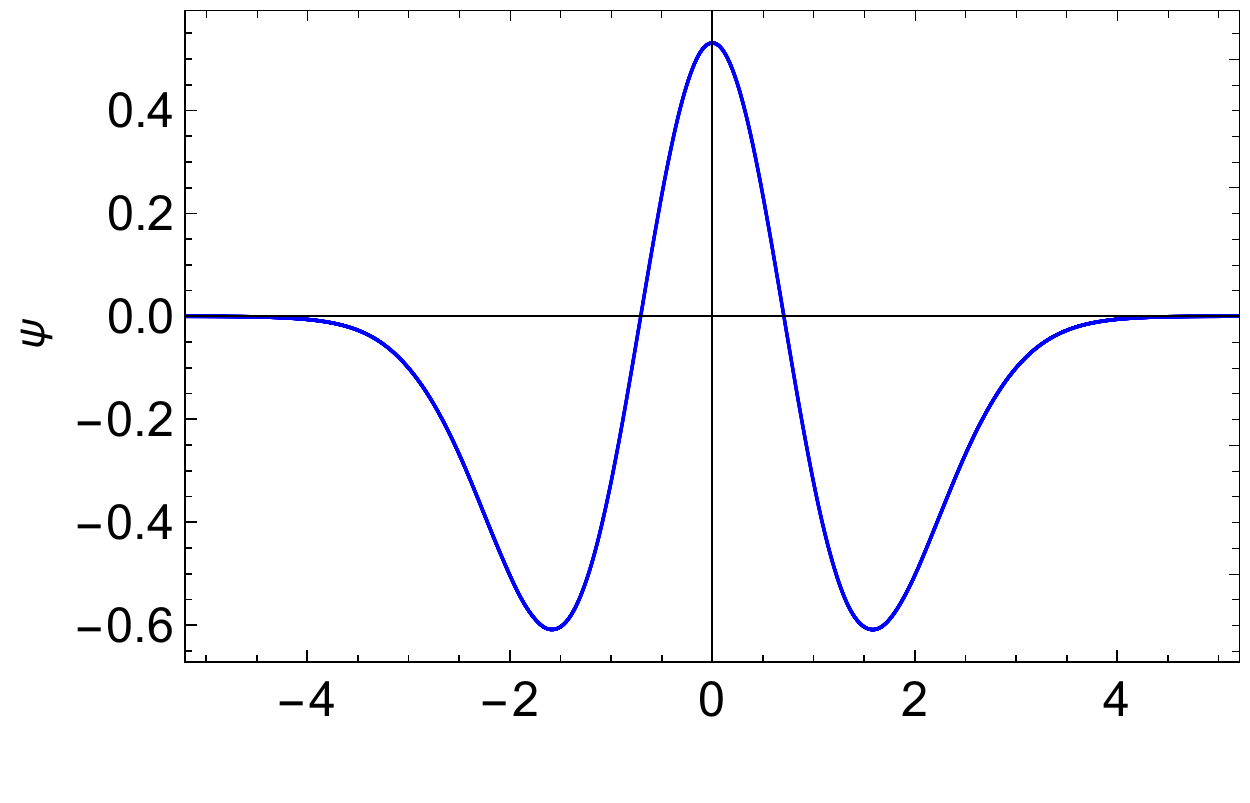} 
     \end{minipage}\hfill
     \begin{minipage}{0.5\columnwidth}
         \centering
         \includegraphics[width=1.0\columnwidth]{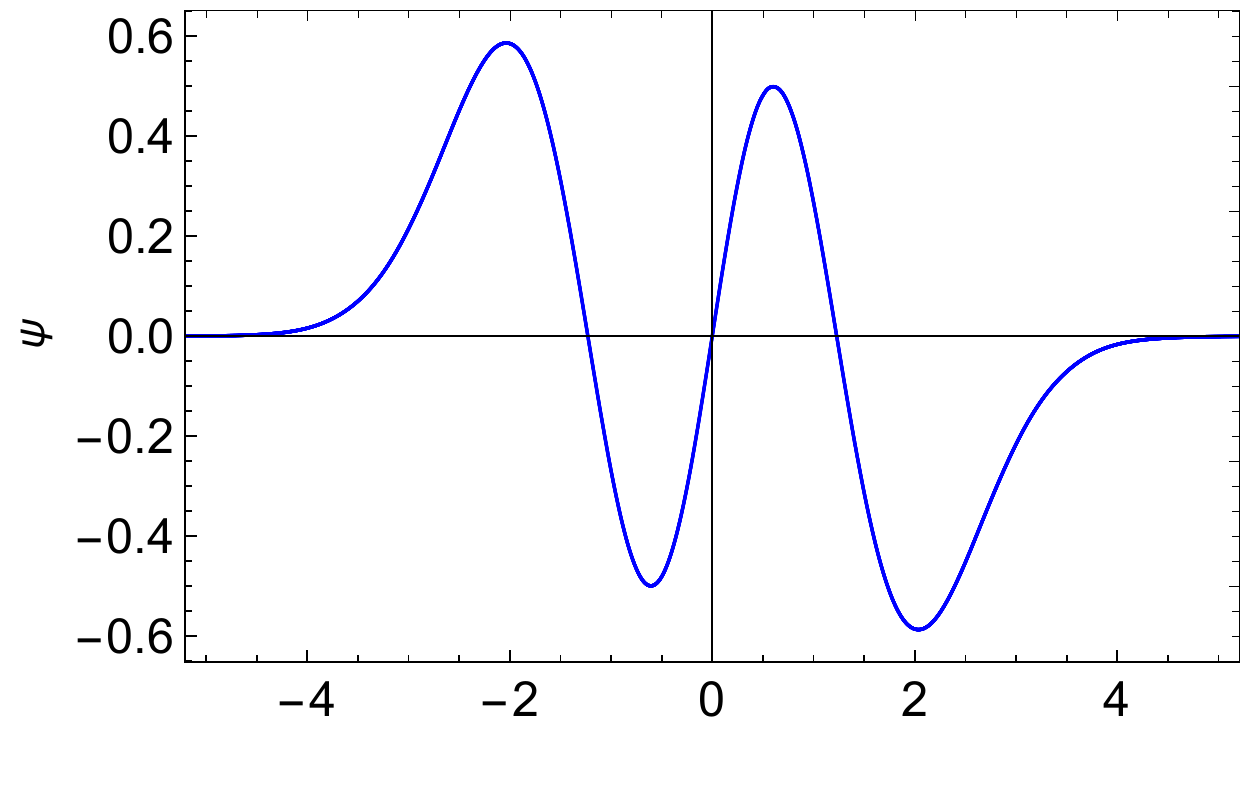} 
     \end{minipage}
     \caption{Predicted $\ket{\psi}$ of one-dimensional harmonic oscillator. Left: second excited state; right: third excited state.}\label{fig:1D23}
 \end{figure}

\begin{figure}[H]
     \centering
     \begin{minipage}{0.5\columnwidth}
         \centering
         \includegraphics[width=1.0\columnwidth]{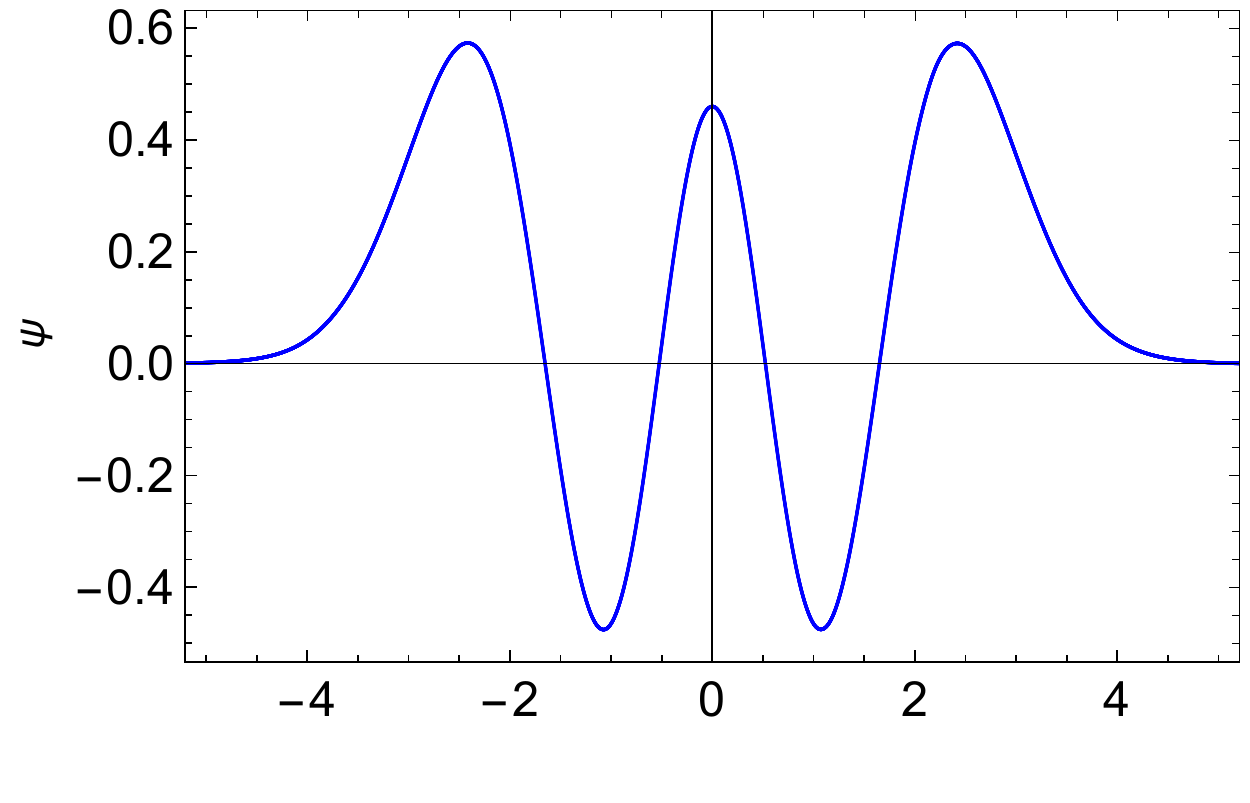} 
     \end{minipage}\hfill
     \begin{minipage}{0.5\columnwidth}
         \centering
         \includegraphics[width=1.0\columnwidth]{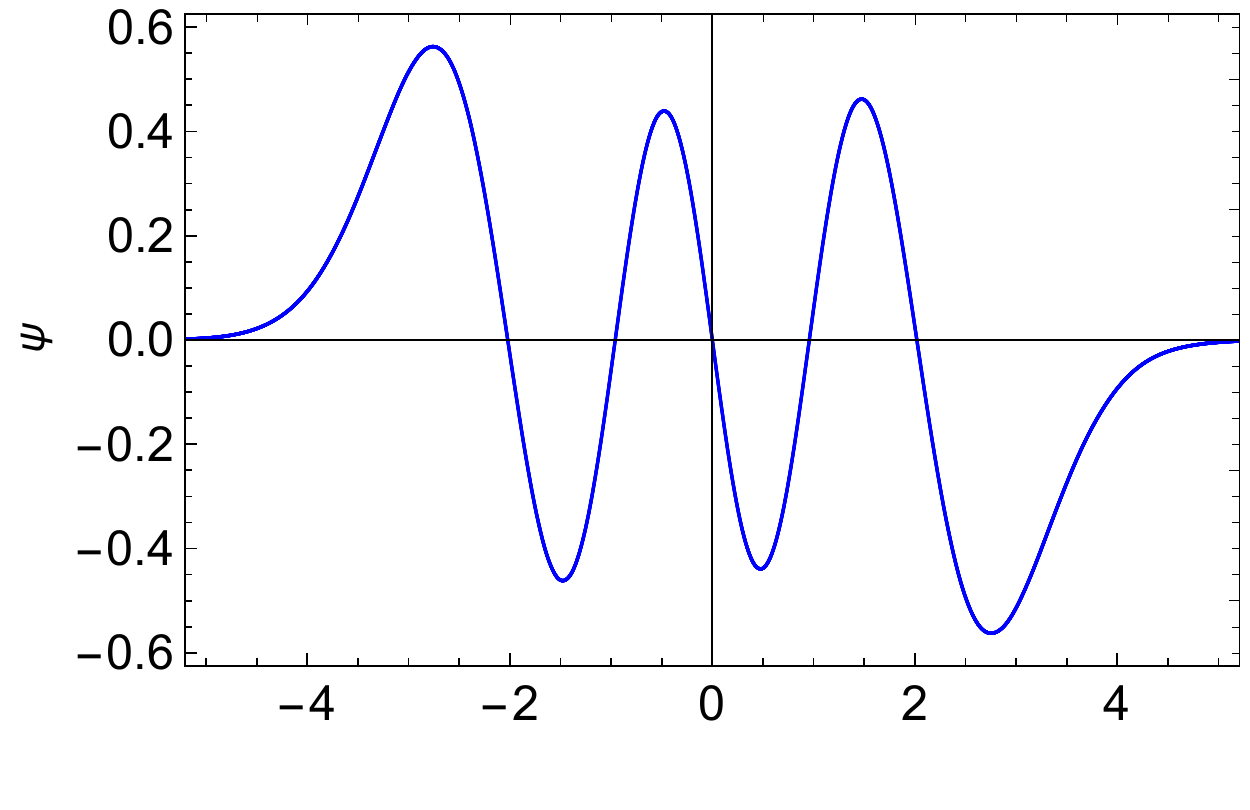} 
    \end{minipage}
     \caption{Predicted $\ket{\psi}$ of one-dimensional harmonic oscillator. Left: $4$-${th}$ excited state; right: $5$-${th}$ excited state.}\label{fig:1D45}
\end{figure}

The corresponding variational energy levels are shown in Table.~\ref{table:1}. 
Fig.~\ref{fig:1D01},~\ref{fig:1D23} and ~\ref{fig:1D45} show the simulated ground state and first to $5$-${th}$ excited states. We use a neural network with 8 hidden layers and each layer has 16 neurons,  $M_x$ = 50,000, $L_x = 10$. The eigenvalues are $(n+\frac{1}{2})$, where $n$ is the quantum number.
Our results agree well with exact results. 
\begin{table}[H]
\small
\caption{Comparison between variational energy levels with analytical energy levels (one-dimensional quantum harmonic oscillator). }
\label{table:1}
\centering
\begin{ruledtabular}
\begin{tabular}{ccccccc}
State             & 0     & 1     & 2    & 3    & 4    & 5    \\ \hline
Simulated $E$  & 0.5000 & 1.500 & 2.500 & 3.502 & 4.500 & 5.501 \\ \hline
Analytical $E$  & 0.5000 & 1.500  & 2.500 & 3.500 & 4.500 & 5.500 \\
\end{tabular}
\end{ruledtabular}
\end{table}
\subsection{Two-dimensional quantum harmonic oscillator}
\begin{figure}[H]
     \centering
     \begin{minipage}{0.5\columnwidth}
         \centering
         \includegraphics[width=1.0\columnwidth]{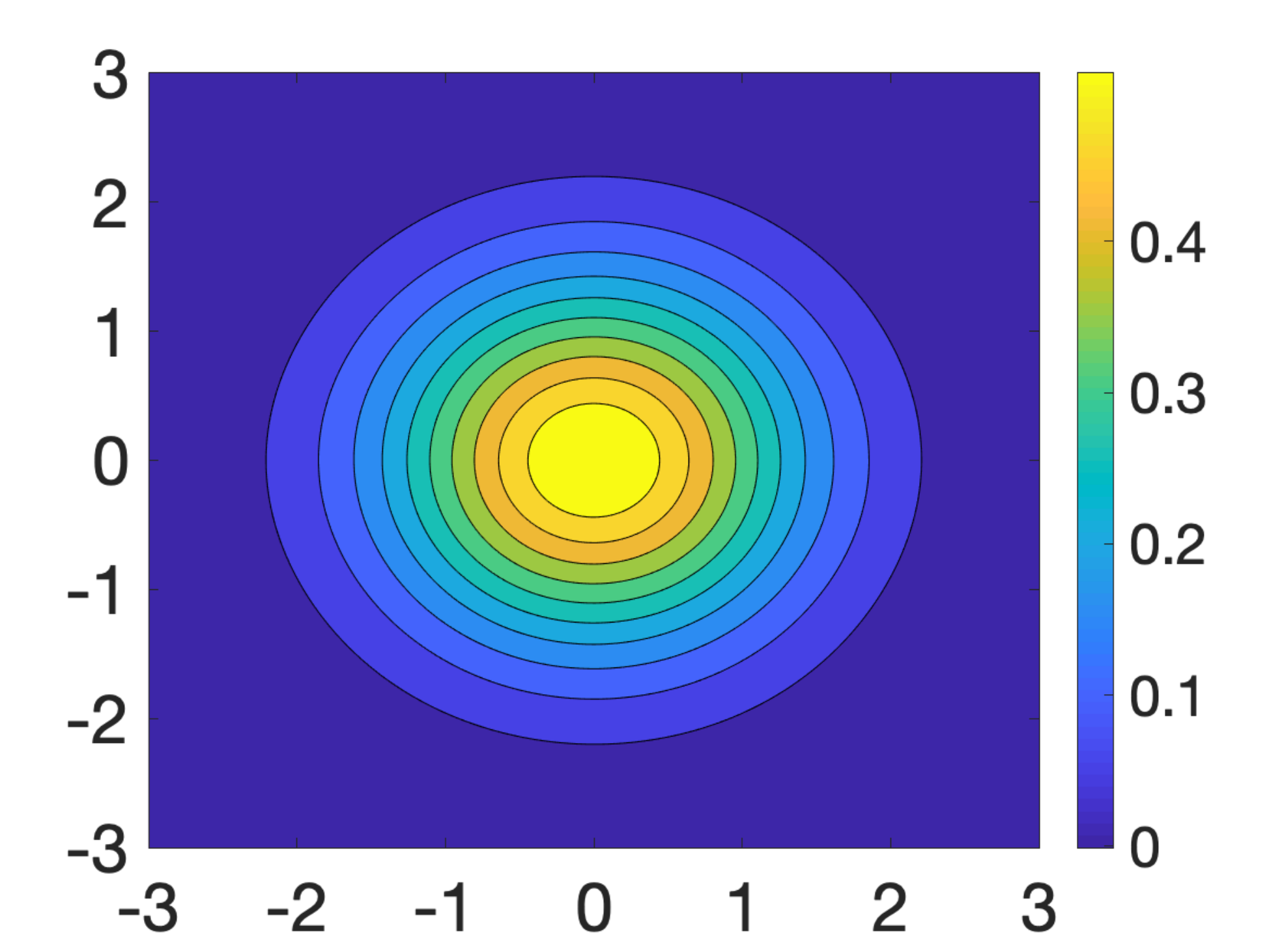} 
     \end{minipage}\hfill
     \begin{minipage}{0.5\columnwidth}
         \centering
         \includegraphics[width=1.0\columnwidth]{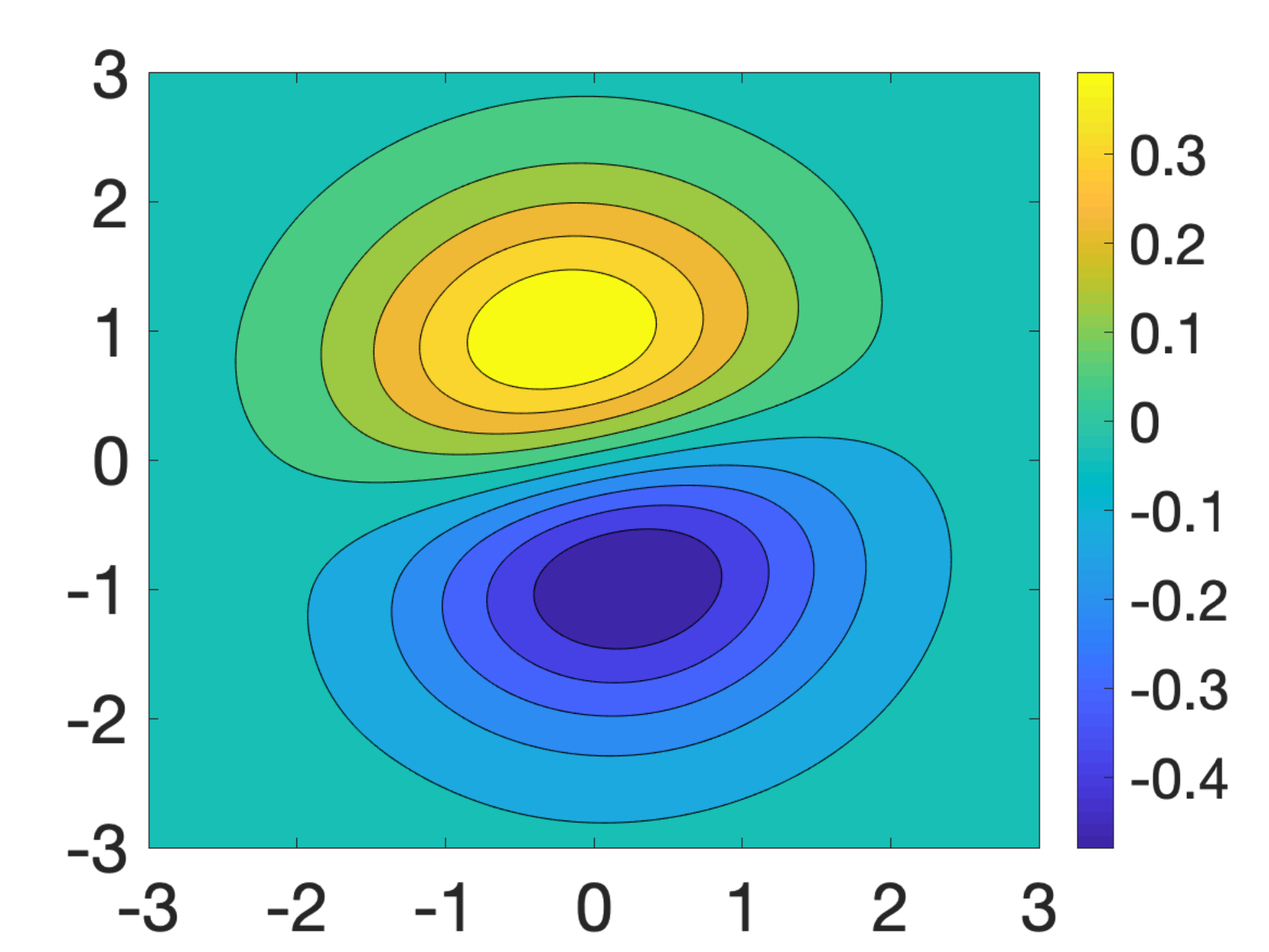} 
     \end{minipage}
     \caption{Predicted $\ket{\psi}$ of two-dimensional harmonic oscillator. Left: ground state; right: first excited state.}\label{fig:2Dgf}
 \end{figure}

\begin{figure}[H]
     \centering
     \begin{minipage}{0.5\columnwidth}
         \centering
         \includegraphics[width=1.0\columnwidth]{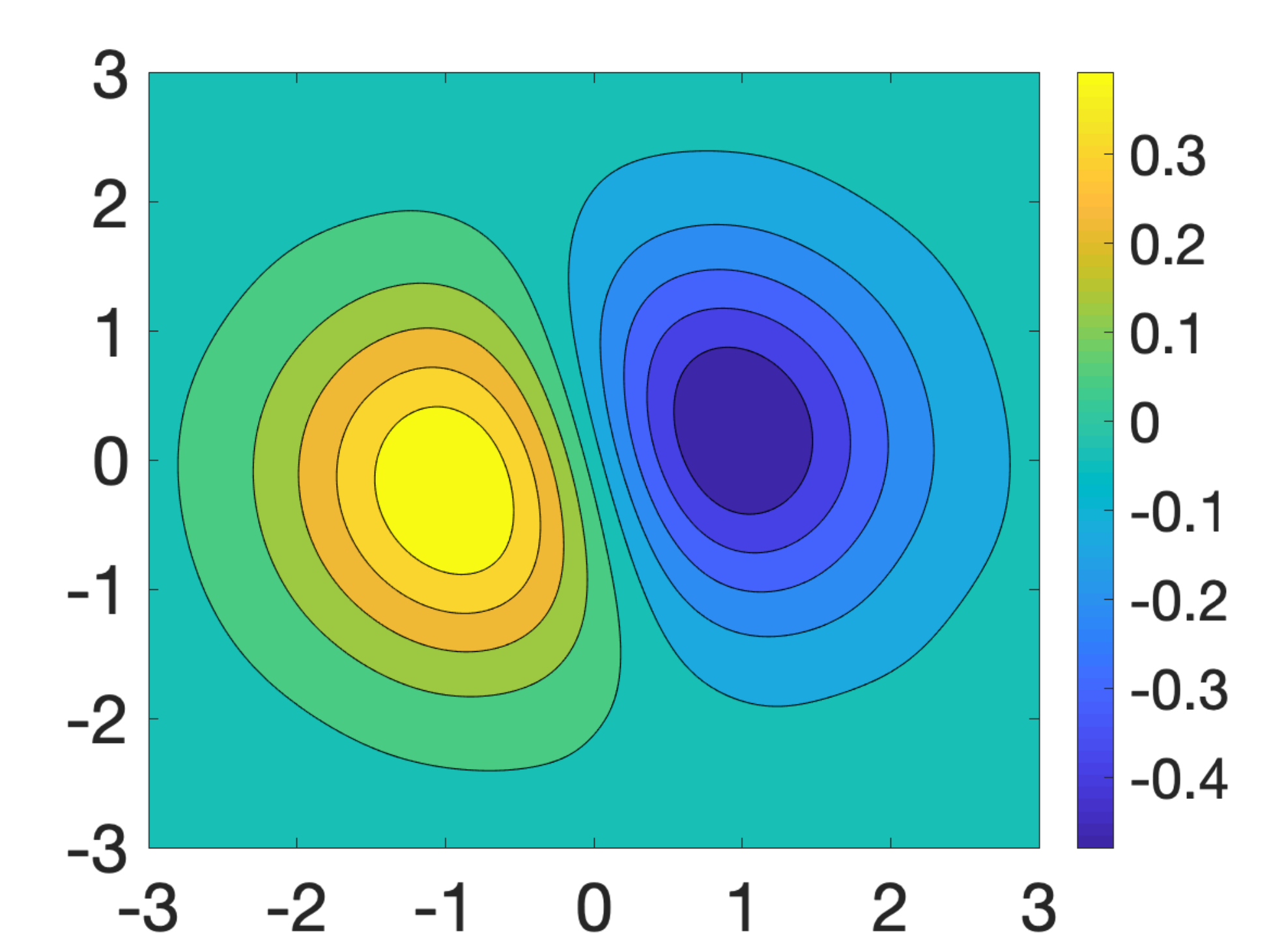} 
     \end{minipage}\hfill
     \begin{minipage}{0.5\columnwidth}
         \centering
         \includegraphics[width=1.0\columnwidth]{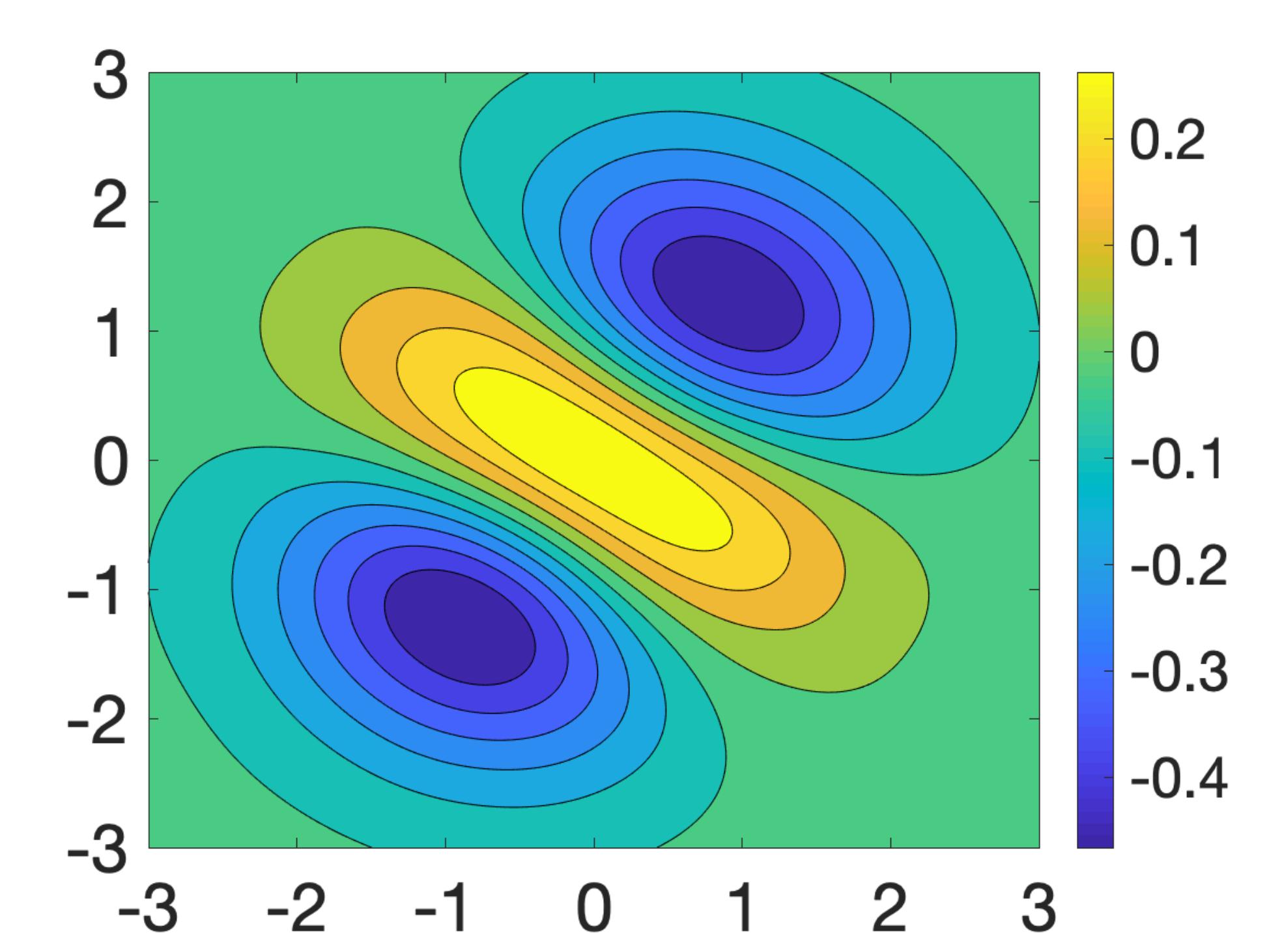} 
     \end{minipage}
     \caption{Predicted $\ket{\psi}$ of two-dimensional harmonic oscillator. Left: second excited state; right: third excited state.}\label{fig:2D23}
 \end{figure}

\begin{figure}[H]
     \centering
     \begin{minipage}{0.5\columnwidth}
         \centering
         \includegraphics[width=1.0\columnwidth]{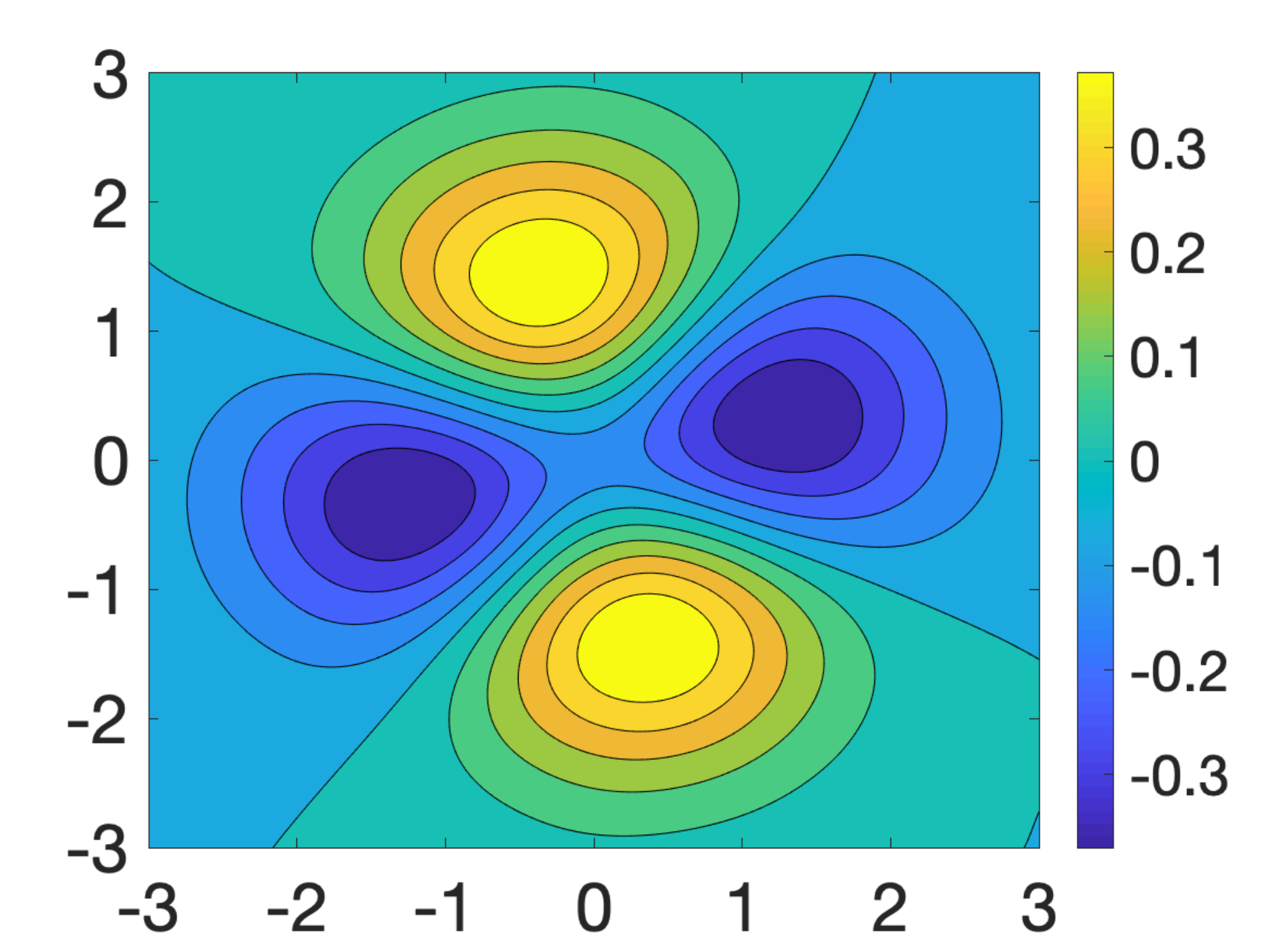} 
     \end{minipage}\hfill
     \begin{minipage}{0.5\columnwidth}
         \centering
         \includegraphics[width=1.0\columnwidth]{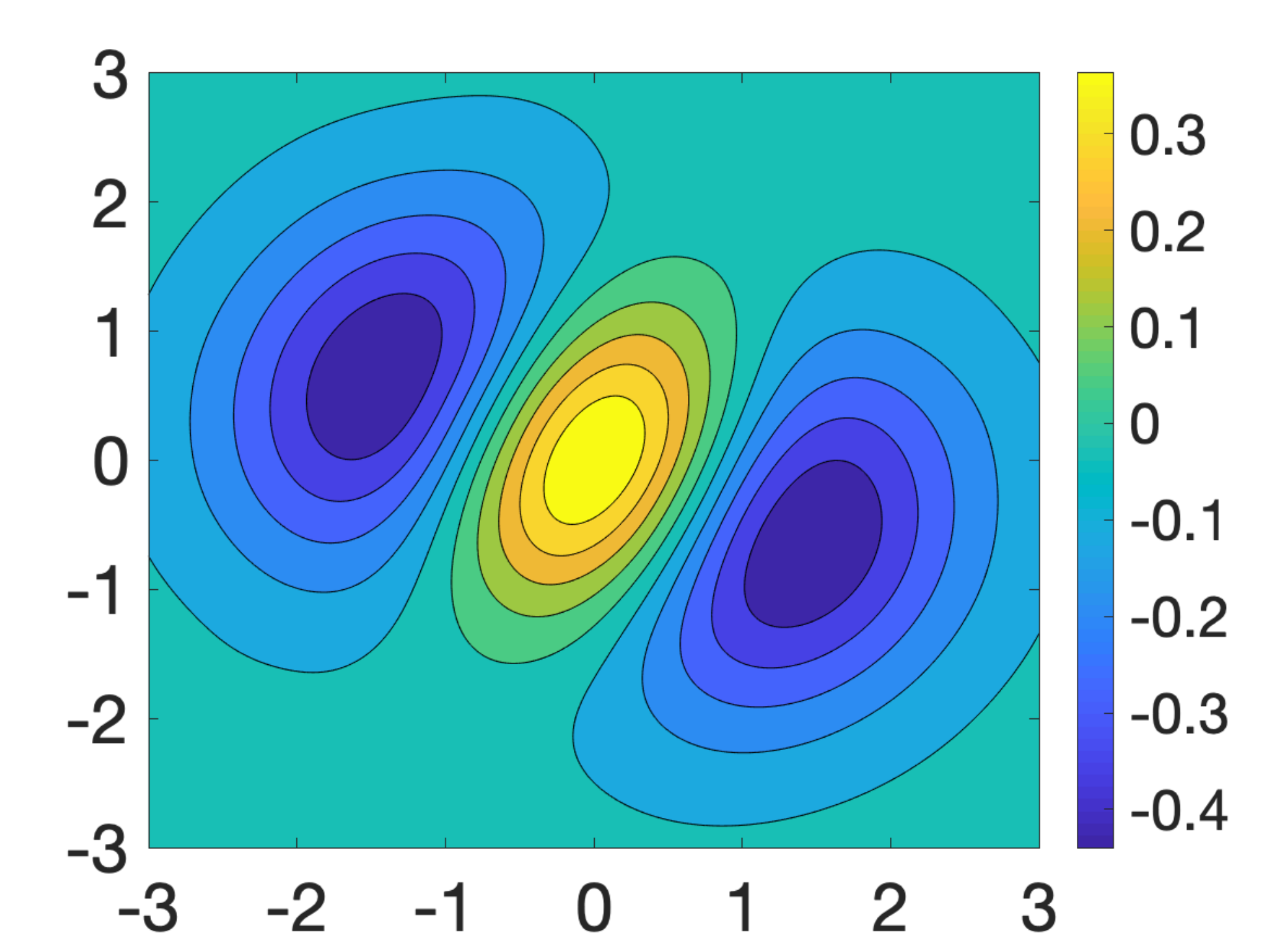} 
     \end{minipage}
     \caption{Predicted $\ket{\psi}$ of two-dimensional harmonic oscillator. Left: 4-$th$ excited state; right:  5-$th$ excited state.}\label{fig:2D45}
 \end{figure}

The predicted variational eigenvalues on two-dimensional quantum harmonic oscillator are shown in  Table~\ref{table:2}, for simplicity, we set $L_x = L_y$ and $M_x = M_y$ during simulation. The eigenvalues are $(n_x + n_y +1)$, where $n_x$ and $n_y$ are the quantum numbers along $x$ and $y$ direction. The results agree with exact solutions.
Figure~\ref{fig:2Dgf},~\ref{fig:2D23} and~\ref{fig:2D45} describe the corresponding wave functions listed in 
Table~\ref{table:2}.
\begin{table}[H]
\caption{Comparison between variational energy levels with analytical energy levels (two-dimensional quantum harmonic oscillator). The neural network has 8 hidden layers and each layer has 16 neurons, $M_x=M_y=500$, $L_x = 10$.}
\label{table:2}
\centering
\begin{ruledtabular}
\begin{tabular}{cccccccc}
State  & 0    & 1    & 2    & 3    & 4  & 5  \\ \hline
Simulated $E$ & 1.000 & 2.000  & 2.001 & 3.001 & 3.001 & 3.001 \\ \hline
Analytical $E$ & 1.000 & 2.000 & 2.000 & 3.000 & 3.000 & 3.000 \\ 
\end{tabular}
\end{ruledtabular}
\end{table}

We further study how the parameters affect the approximation accuracy using a neural network wave function ansatz. Let  $\Delta E = |E_{var} - E_{ana}|$ denote the approximation error  between variational energy $E_{var}$ and analytical solution $E_{ana}$. As discussed before, since the evaluation of the Hamiltonian depends on the input coordinates, we anticipate that a denser discretization will result in improved accuracy. We here study relationship between  $\Delta E$ and mesh size $M_x$. \begin{figure}[H]
     \centering
     \begin{minipage}{0.5\columnwidth}
         \centering
         \includegraphics[width=1.0\columnwidth]{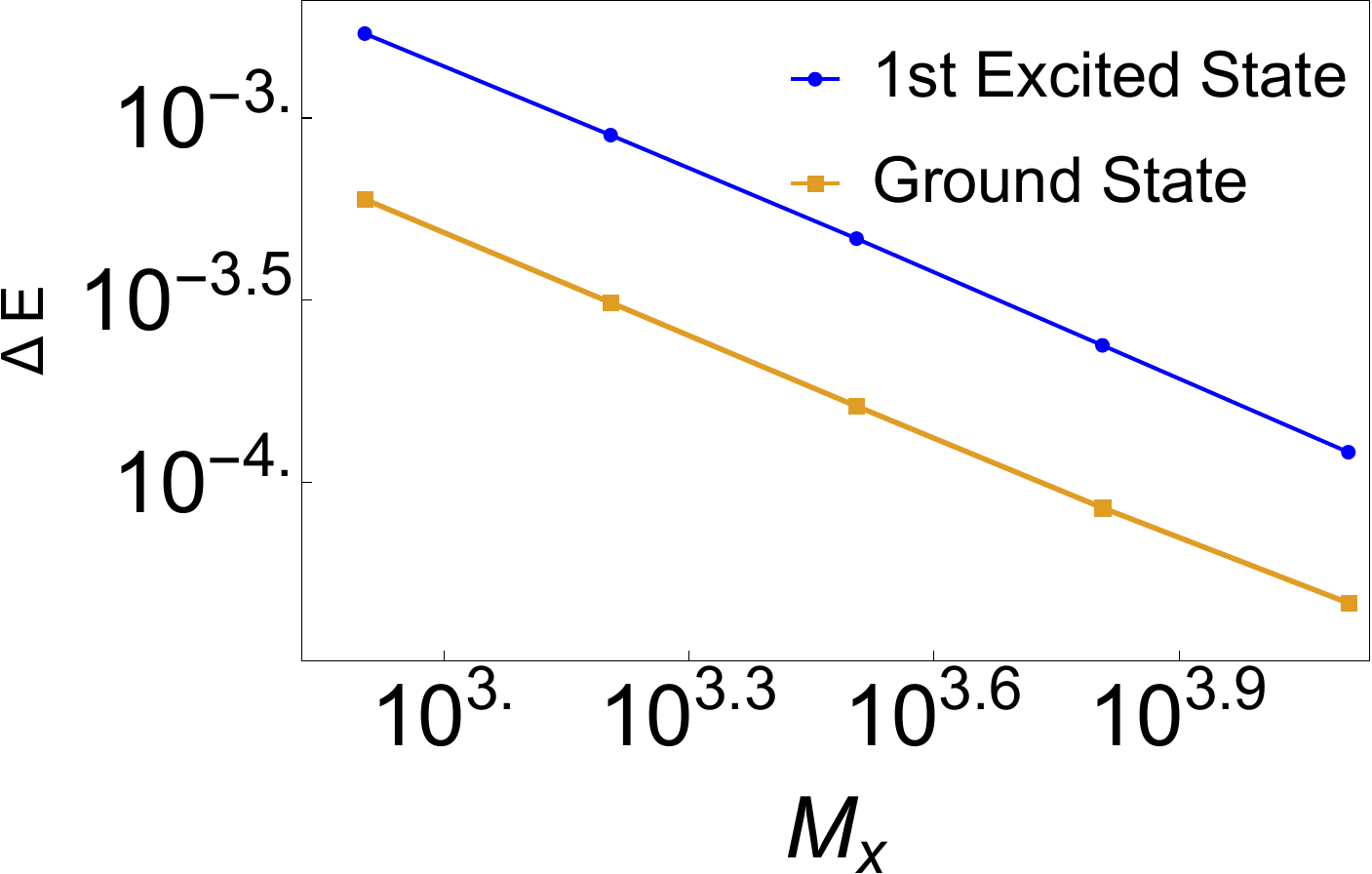} 
     \end{minipage}\hfill
     \begin{minipage}{0.5\columnwidth}
         \centering
         \includegraphics[width=1.0\columnwidth]{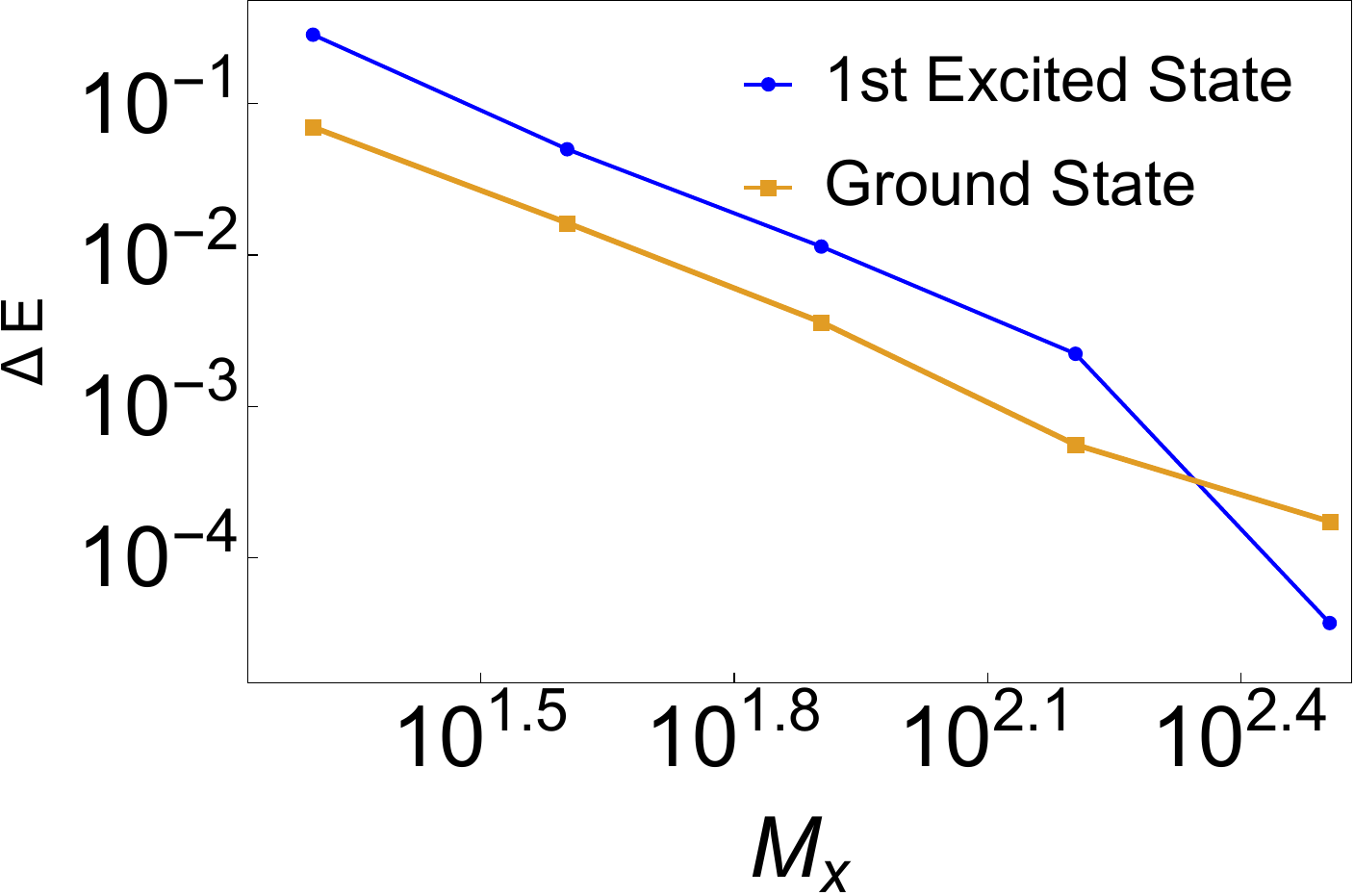} 
    \end{minipage}
     \caption{Approximation error with discretization parameter $M_x$. Left: 1D harmonic oscillator ($L_x = 10, \sigma=512$, $l=1$); Right: 2D harmonic oscillator ($L_x = 10, \sigma=2048$, $l=1$). }\label{fig:delE}
\end{figure}
We use a neural network  with only one hidden layer. Although we are using a simple set up, our neural network ansatz has enough expressivity for any wave function, given that one hidden layer with sufficient hidden units is capable of approximating any wave function \cite{hornik1989multilayer}.

Figure~\ref{fig:delE} illustrates how $\Delta E$ depends on $M_x$. 
We increase $M_x$ in powers of 2 and we observe a linear trend between $\Delta E$ and $M_x$ from the log-log plot.
Our results suggest that $\Delta E$ decreases as mesh size $M_x$ increases.
We refer more error analysis to  supplementary material.

\section{Conclusion~\label{sec:3}}
In this letter, we demonstrate the success of a variational approach, based on a neural network wave function ansatz, for approximating the single-particle ground state and  excited states.
Our method enforces orthogonality to previously converged states, which leads to the rearrangement of eigenvalues.
We achieve high accuracy in the description of excited states. 
This method can be generalized to any Hermitian operator.

\bibliography{arxiv}
\section{Supplementary Material~\label{sec:4}}
\beginsupplement
\subsection{Computational details}
Code repository: \href{https://github.com/yimengmin/vmc}{https://github.com/yimengmin/vmc}.

We train our model for 200,000 steps using Adam optimizer \cite{kingma2014adam}. The initial learning rate is 0.01 and we decrease the learning rate by a factor of 0.9 every 20,000 steps.  
\subsection{Error analysis}
\begin{figure}
     \centering
     \begin{minipage}{0.48\columnwidth}
         \centering
         \includegraphics[width=1.0\columnwidth]{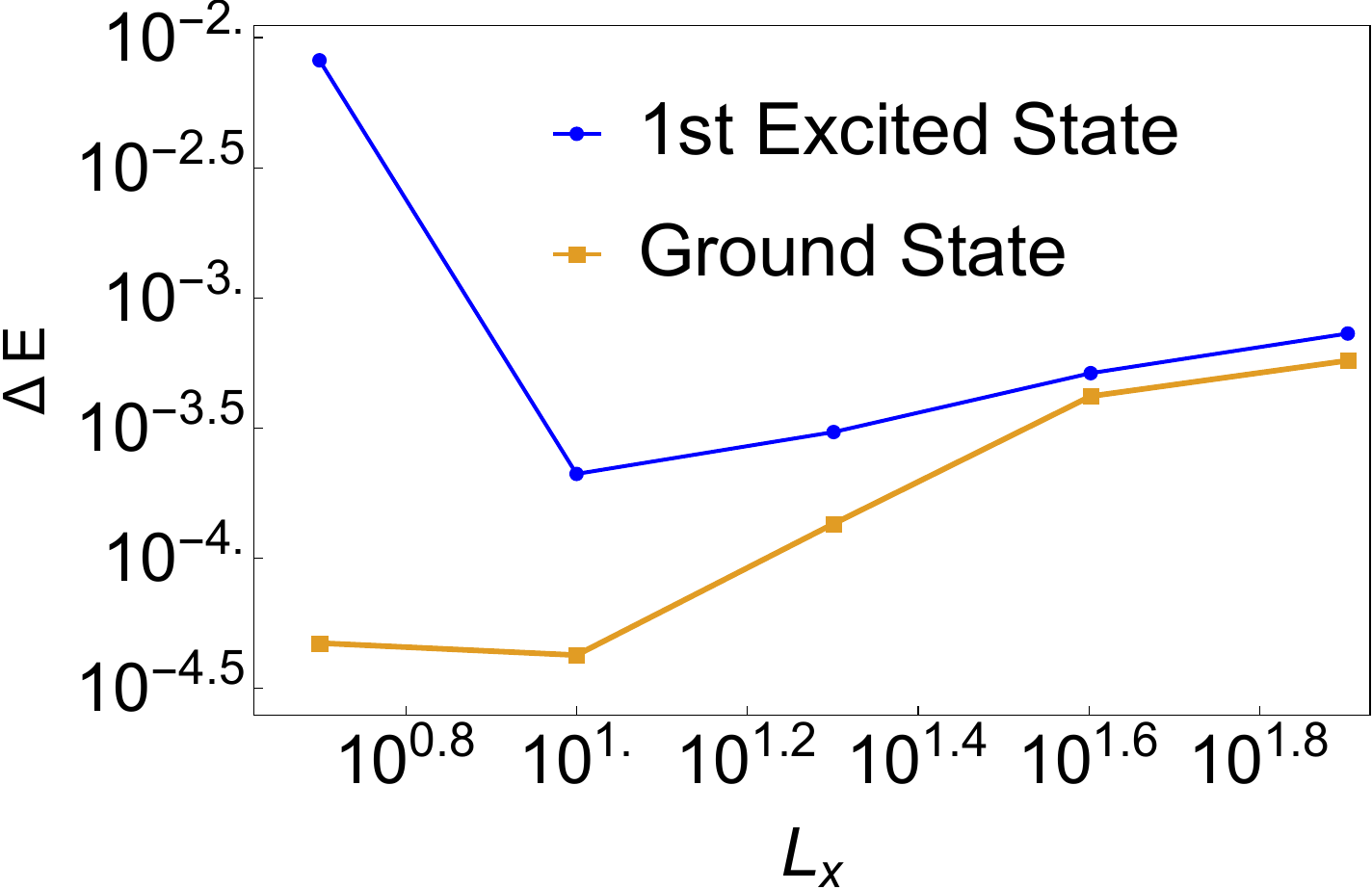} 
     \end{minipage}\hfill
     \begin{minipage}{0.48\columnwidth}
         \centering
         \includegraphics[width=1.0\columnwidth]{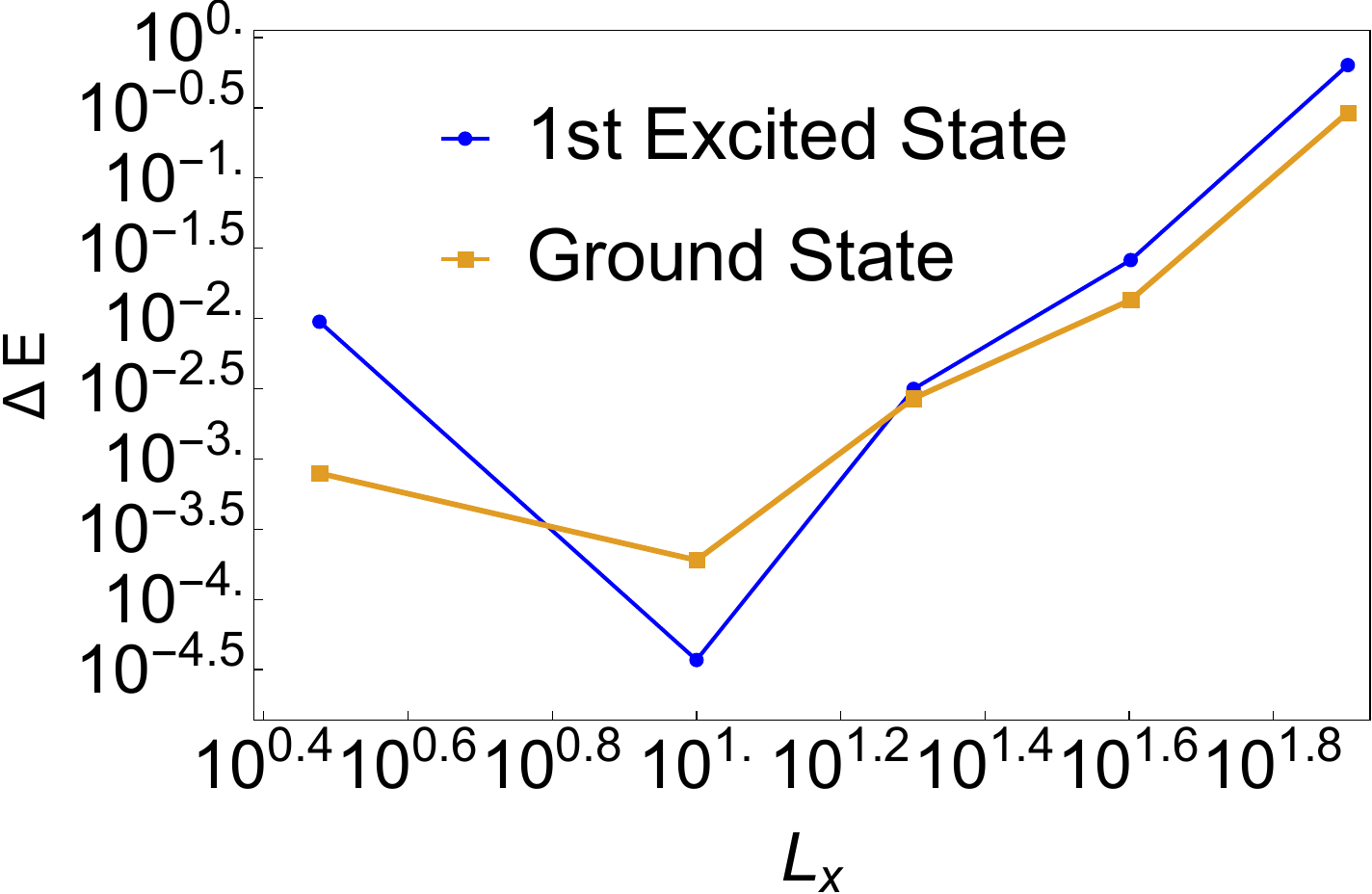} 
    \end{minipage}
     \caption{Approximation error with  scale parameter $L_x$. Left: 1D harmonic oscillator ($M_x = 12800$, $\sigma=512$, $l=1$); Right: 2D harmonic oscillator ($M_x = 320$, $\sigma=2048$, $l=1$). }\label{fig:delEscale}
\end{figure}

\begin{figure}
     \centering
     \begin{minipage}{0.48\columnwidth}
         \centering
         \includegraphics[width=1.0\columnwidth]{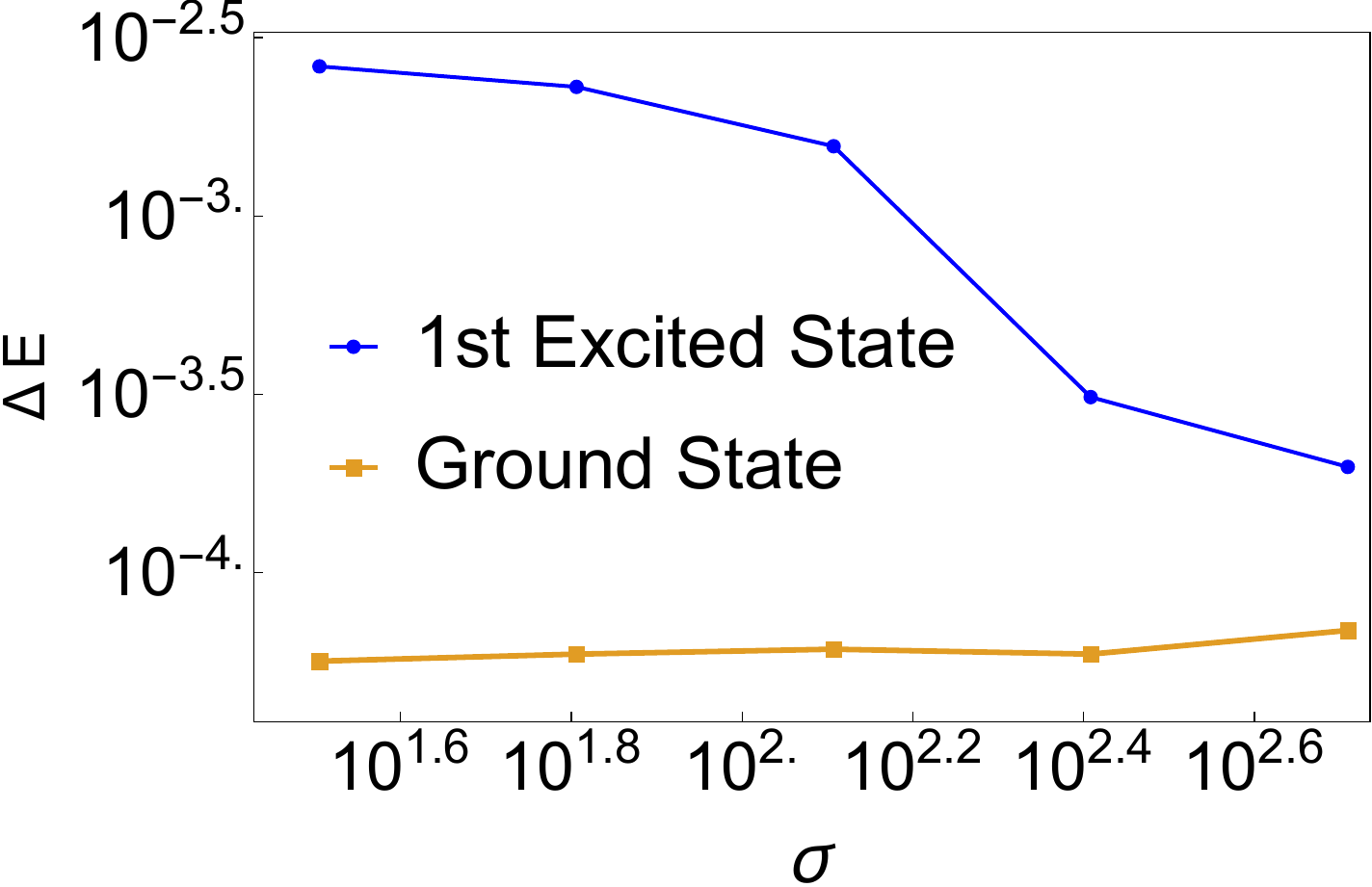} 
     \end{minipage}\hfill
     \begin{minipage}{0.48\columnwidth}
         \centering
         \includegraphics[width=1.0\columnwidth]{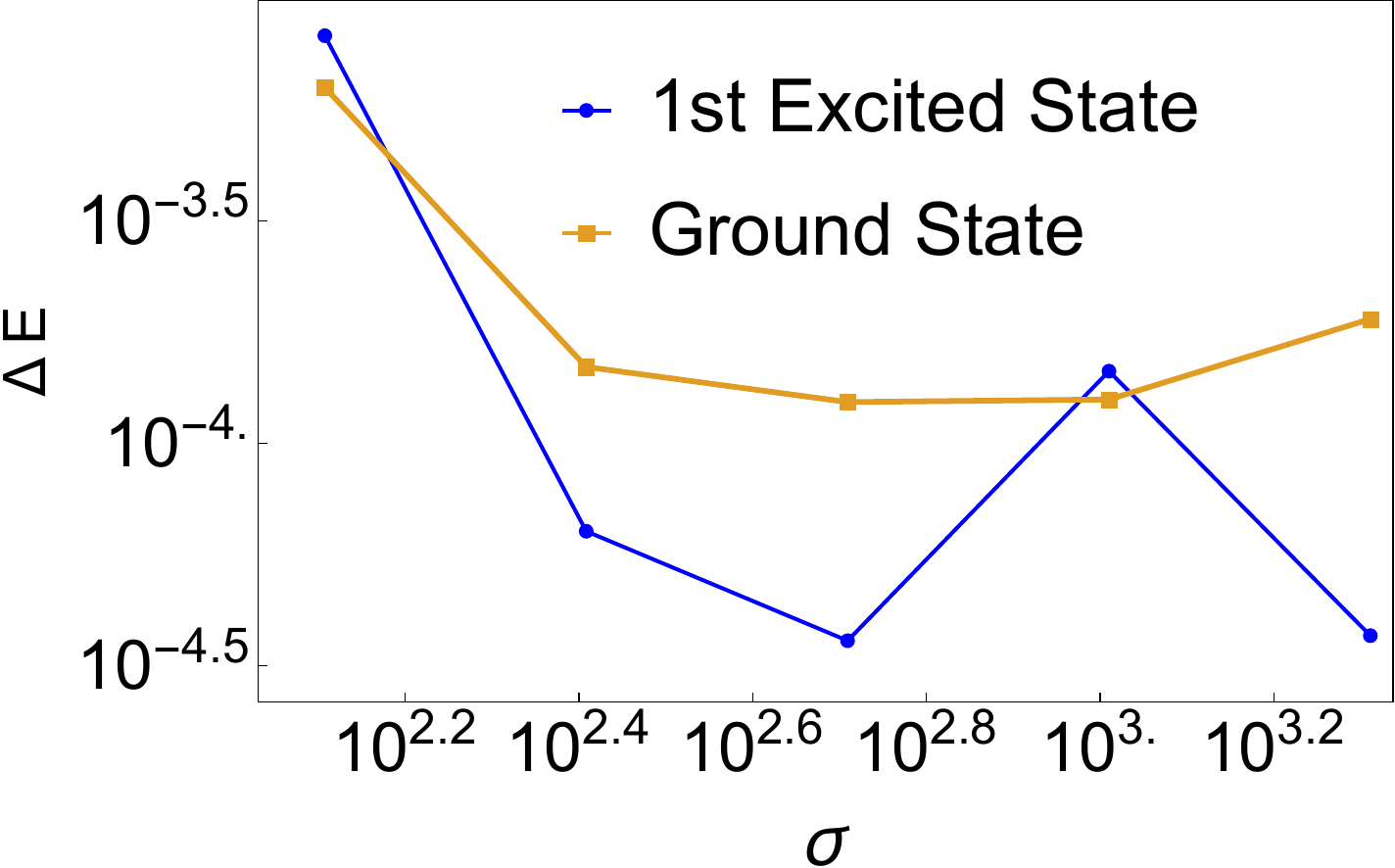} 
    \end{minipage}
     \caption{Approximation error with width parameter $\sigma$. Left: 1D harmonic oscillator ($L_x = 10, M_x = 12800$, $l=1$); Right: 2D harmonic oscillator ($L_x = 10, M_x = 320$,  $l=1$). }\label{fig:delEwidth}
\end{figure}

Figure~\ref{fig:delEscale} demonstrates how $L_x$ affects the approximation accuracy. 
Our results suggest there exists an optimal scale.  Small $L_x$ is not capable to provide enough support for the wave function, while a larger $L_x$ corresponds to the reduction of $M_x$, which also weakens the simulation accuracy.
\begin{figure}
     \centering
     \begin{minipage}{0.48\columnwidth}
         \centering
         \includegraphics[width=1.0\columnwidth]{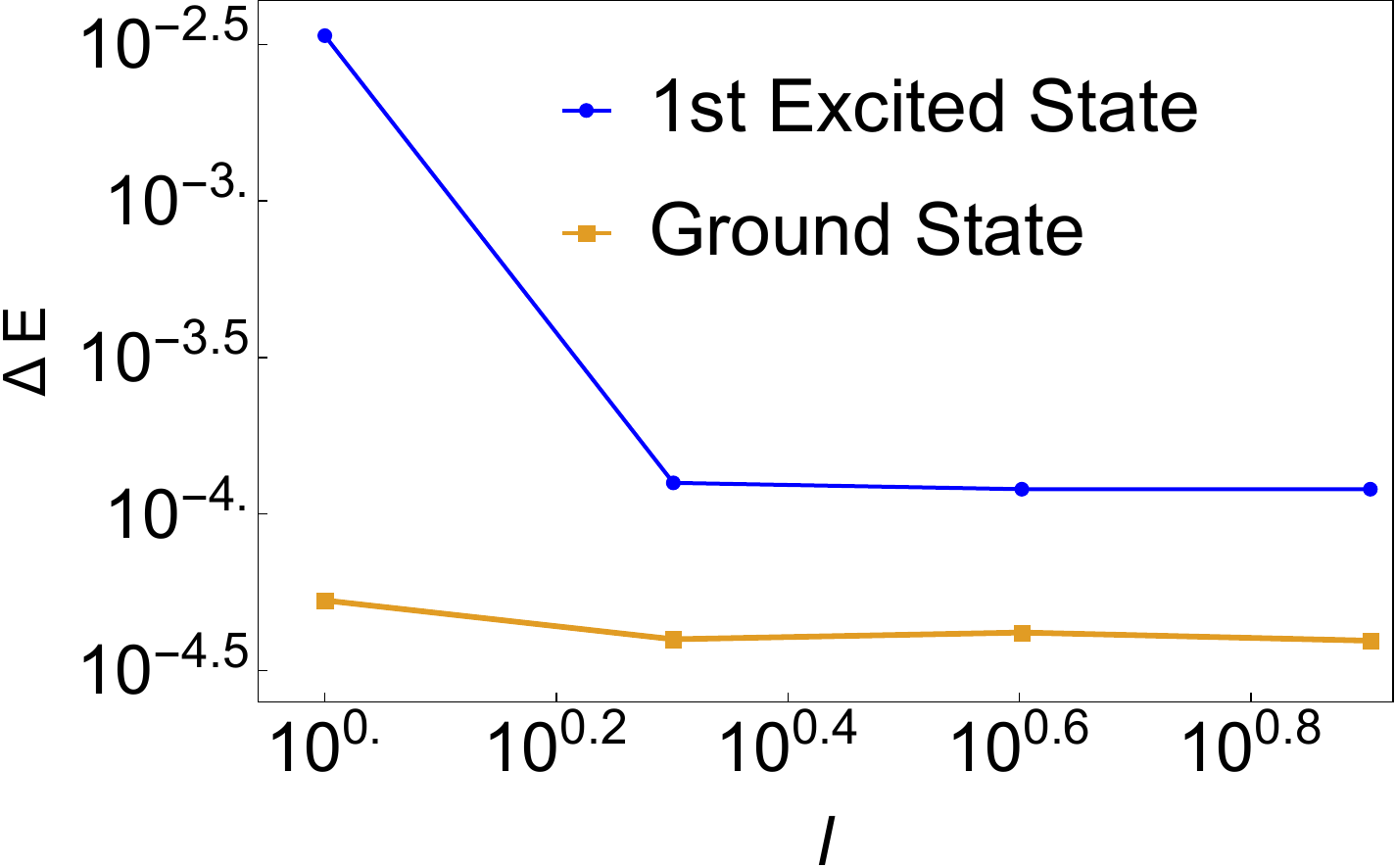}
     \end{minipage}\hfill
     \begin{minipage}{0.48\columnwidth}
         \centering
         \includegraphics[width=1.0\columnwidth]{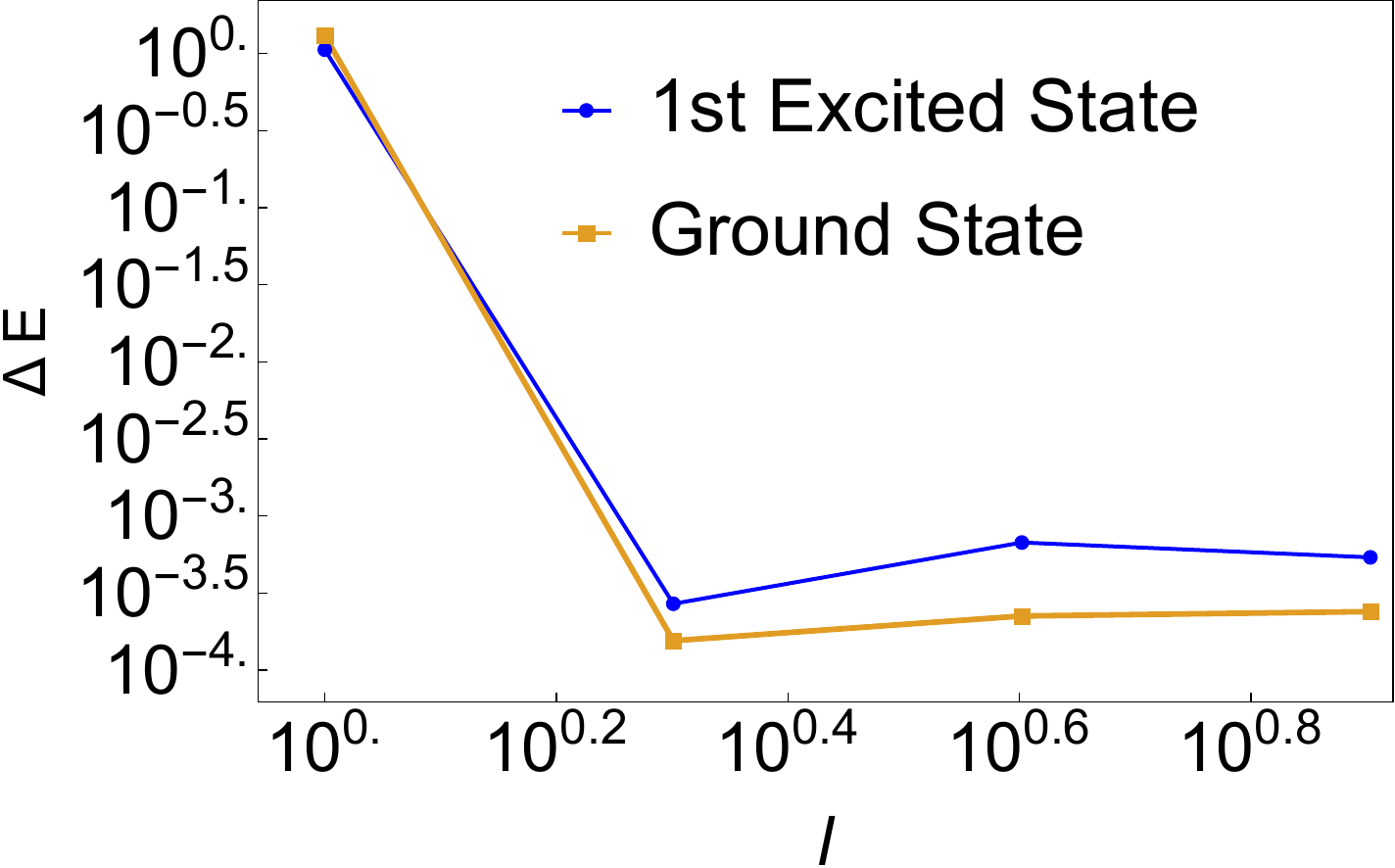}
    \end{minipage}
     \caption{Approximation error with  depth parameter $l$. Left: 1D harmonic oscillator ($M_x = 12800$, $\sigma=16$, $L_x = 10$); Right: 2D harmonic oscillator ($M_x = 320$, $\sigma=16$, $L_x = 10$). }\label{fig:delEdepth}
\end{figure}
Figure~\ref{fig:delEdepth} shows how   layer depth parameter $l$ affects $\Delta E$. As $l$ increases, $\Delta E$ decreases first and then maintains at the same level. We observe a slight increase when we  further deepen the network in 2D condition, suggesting that a very deep neural network is not necessary.
Figure~\ref{fig:delEwidth} shows how the width parameter $\sigma$ affects $\Delta E$. Our results suggest that a very wide neural network is also unnecessary.


\subsection{Comparison of number of parameters}
The NN-based variational method has fewer parameters comparing to data-driven methods. For example, Mills $et~al.$ study the two-dimensional single electron system using a convolutional deep neural network \cite{mills2017deep} with error $ \sim 1 e^{-3}$. The convolutional deep neural network has seven layers and the final convolutional layer is fed into a fully connected
layer of width 1024, this operation creates $4\times4\times16\times1024 = 262,144$ parameters. In our simulation, we use a single-layer neural network with 2048 hidden units for Figure~\ref{fig:delE} and the total number of parameters is 6144. To further reduce the number of parameters, 
we use 8 hidden layers and each hidden layer has a width of 16, then the total number of parameters for two-dimensional case is 2096 (Figure~\ref{fig:2Dgf},~\ref{fig:2D23},~\ref{fig:2D45}).
Our data-free method has fewer parameters than the data-driven counterpart.

\end{document}